\documentclass{Interspeech}

\interspeechcameraready 

\title{Tackling Cognitive Impairment Detection from Speech:\\A submission to the PROCESS Challenge}

\author[affiliation={1}]{Catarina}{Botelho$\ddagger$}
\author[affiliation={3}]{David}{Gimeno-Gómez$\ddagger$}
\author[affiliation={1}]{Francisco}{Teixeira$\ddagger$}
\author[affiliation={1,2}]{John}{Mendonça}
\author[affiliation={1,2}]{Patrícia}{Pereira}
\author[affiliation={1,2}]{Diogo A.P.}{Nunes}
\author[affiliation={1}]{Thomas}{Rolland}
\author[affiliation={1}]{Anna}{Pompili}
\author[affiliation={1}]{Rubén}{Solera-Ureña}
\author[affiliation={1,2}]{Maria}{Ponte}
\author[affiliation={1,2}]{David}{Martins de Matos}
\author[affiliation={3}]{Carlos-D.}{Martínez-Hinarejos}
\author[affiliation={1,2}]{Isabel}{Trancoso}
\author[affiliation={1,2}]{Alberto}{Abad}

\affiliation{INESC-ID, Portugal; $^2$Instituto Superior Técnico}{University of Lisbon}{Portugal}
\affiliation{PRHLT research center}{Universitat Politècnica de València}{Spain}
\email{catarina.t.botelho@inesc-id.pt}

\usepackage{stfloats}
\usepackage{booktabs}
\usepackage{multicol}
\usepackage{arydshln}
\usepackage{multirow}
\usepackage{arydshln}
\usepackage{adjustbox}
\usepackage{threeparttable}
\usepackage[caption=false,font=tiny,labelfont=sf,textfont=sf]{subfig}
\usepackage{comment}
\usepackage{array}
\usepackage{url}

\begin{document}

\maketitle
\vspace*{10pt}
\begin{abstract}
This work describes our group's submission to the PROCESS Challenge 2024, with the goal of assessing cognitive decline through spontaneous speech, using three guided clinical tasks.
This joint effort followed a holistic approach, encompassing both knowledge-based acoustic and text-based feature sets, as well as LLM-based macrolinguistic descriptors, pause-based acoustic biomarkers, and multiple neural representations (e.g., LongFormer, ECAPA-TDNN, and Trillson embeddings).
Combining these feature sets with different classifiers resulted in a large pool of models, from which we selected those that provided the best balance between train, development, and individual class performance.
Our results show that our best performing systems correspond to combinations of models that are complementary to each other, relying on acoustic and textual information from all three clinical tasks.
\end{abstract}

\section{Introduction}
\label{sec:intro}

Dementia, which affects fifty-five million people around the world, is one of the major causes of disability and dependency among older people~\cite{who-dementia}. It is marked by a progressive decline in cognitive functions beyond what is considered normal in biological aging~\cite{who-dementia}.
Alzheimer's disease (AD), a progressive neurodegenerative disorder, is the most common form of dementia. Although memory impairment is the most prominent symptom, speech and language impairments are also persistent and useful in detecting early onset of the disease~\cite{snowdon1996linguistic, kempler1995, reilly2011}.
Mild cognitive impairment (MCI) is a clinical condition that often leads to dementia. It is characterized by noticeable deficits in one or more cognitive domains that exceed typical age-related changes but do not meet the diagnostic criteria for dementia~\cite{petersen2001current, who2017}.

Although numerous studies have leveraged speech and language biomarkers for detecting dementia, AD, and MCI (e.g., \cite{satt2014speech, bertini2021automatic, perez2024multilingual, pompili2020pragmatic, ablimit_botelho2022exploring, luz2021alzheimer}), the field remains hindered by a lack of standardization and sufficiently large, representative benchmarks for systematic comparison of methodologies. Notable exceptions include the ADReSS challenges, which focus on AD detection (e.g., \cite{ADReSS, ADReSSo_luz2021detecting, ADReSSM_luz2023multilingual}), and the Taukadial challenge, which incorporates data in English and Mandarin to address MCI detection (e.g., \cite{TAUKADIAL_luz2024connected}). Recently, the PROCESS challenge introduced a dataset designed for three-class classification, distinguishing healthy controls, individuals with MCI, and patients with dementia, which is less commonly studied \cite{christensen2024process}.
\vspace{-1pt}
\let\thefootnote\relax\footnote{$\ddagger$ Equal contributions.}

\begin{figure}[ht]
\vspace{25pt}
    \centering
    \includegraphics[width=1.0\columnwidth]{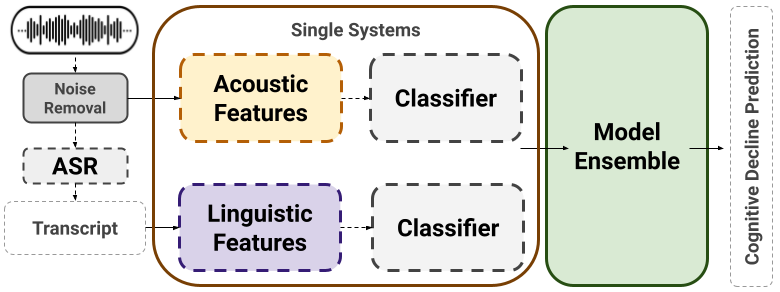}
    \caption{Overall schema of our proposed method.}
    \label{fig:method}
    \vspace{-10pt}
\end{figure}

This study is the result of a collaboration between researchers with diverse expertise in speech and language with the goal of participating in the PROCESS challenge 2024. Therefore, it compares very diverse approaches for the three-class classification task, exploring various data representations, including knowledge-based (KB) acoustic and linguistic features, text- and speech-based neural embeddings, and macrodescriptors derived from LLMs, combined with multiple classification strategies. The resulting top-performing systems were fused using logistic regression to achieve two final systems. This fusion strategy allowed us to leverage complementary information and substantially improve the final classification performance. The diagram in Figure~\ref{fig:method} systematizes our approach.

\section{Corpus}
\label{sec:data}

The PROCESS Challenge dataset \cite{christensen2024process} is designed to promote research on early detection of dementia through speech analysis, emphasizing practical applications in real-world scenarios.

\noindent\textbf{Dataset Overview.} The corpus includes three diagnostic classes, representing a scenario for diagnosing early-stage dementia. The \textbf{Healthy Control (HC)} group comprises volunteers without diagnosed cognitive impairment, as well as individuals who may experience memory issues that are not related to neurodegenerative conditions. The dataset also includes subjects with \textbf{Mild Cognitive Impairment (MCI)}, as well as subjects already diagnosed with early-stage \textbf{Dementia}, with varying degrees of cognitive decline. Detailed demographic and disease severity statistics for the official training and development sets of the challenge are reported in Table~\ref{tab:demographics}.

\begin{table}[t]
\centering
\scriptsize
\caption{Demographic and severity statistics of the official training and development sets of the PROCESS 2024 Challenge.}
\label{tab:demographics}

\begin{adjustbox}{max width=\columnwidth}
\begin{threeparttable}
\begin{tabular}{lccccccc} 
 \toprule
 
\multirow{2}{*}{\textbf{Group}} & \multirow{2}{*}{\textbf{Gender}} & \multicolumn{2}{c}{\textbf{\#Subjects}} & \multicolumn{2}{c}{\textbf{Age}} & \multicolumn{2}{c}{\textbf{MMSE Score}\tnote{$\ddagger$}} \\ \cmidrule{3-8}

 & & \textit{train} & \textit{dev} & \textit{train} & \textit{dev} & \textit{train} & \textit{dev}  \\ \midrule
 
 \multirow{2}{*}[0pt]{\textbf{HC}}\tnote{$\dagger$} 
 
 & \textit{Male} & 23 & 12 & 64.8{\tiny $\pm$13.3} & 63.6{\tiny $\pm$13.1} & 29.0{\tiny $\pm$1.0} & 29.2{\tiny $\pm$1.2} \\
 & \textit{Female} & 37 & 9 & 62.9{\tiny $\pm$12.1} & 61.0{\tiny $\pm$14.3} & 28.9{\tiny $\pm$0.7} & 29.5{\tiny $\pm$0.5} \\ \midrule

 \multirow{2}{*}[0pt]{\textbf{MCI}} 
 
 & \textit{Male} & 22 & 7 & 69.1{\tiny $\pm$7.9} & 68.1{\tiny $\pm$10.8} & 27.1{\tiny $\pm$1.8} & 23.7{\tiny $\pm$3.4} \\
 & \textit{Female} & 22 & 8 & 69.8{\tiny $\pm$10.3} & 61.6{\tiny $\pm$13.0} & 26.5{\tiny $\pm$2.6} &  27.0{\tiny $\pm$1.6} \\ \midrule

 \multirow{2}{*}[0pt]{\textbf{Dementia}} 
 
 & \textit{Male} & 8 & 3 & 75.8{\tiny $\pm$8.0} & 67.0{\tiny $\pm$3.7} & 26.3{\tiny $\pm$2.1} & 27.7{\tiny $\pm$0.9} \\
 & \textit{Female} & 4 & 1 & 69.0{\tiny $\pm$6.3} & 60.0{\tiny $\pm$0.0} & 20.0{\tiny $\pm$0.0} & -- \\

 \bottomrule

\end{tabular}

\begin{tablenotes}
    \scriptsize

    \item[$\dagger$] Gender was not reported for 1 subject but included in our experiments.
        
    \item[$\ddagger$] The MMSE score was only provided for a subset of 69 subjects.

\end{tablenotes}

\end{threeparttable}
\end{adjustbox}

\vspace{-12pt}
\end{table}

\noindent\textbf{Cognitive Assessment Scores.} In addition to the discrete classification labels, the dataset also offers the Mini-Mental State Exam (MMSE) score for a subset of participants. This score is often used to measure cognitive impairment in clinical settings.

\noindent\textbf{Cognitive Assessment Tasks.} Based on neuroscience research for dementia diagnosis, the corpus includes audio recordings from three types of elicitation tasks:

\begin{itemize}

    \item \textbf{Cookie Theft Description (CTD).} Subjects describe the well-known ``Cookie Theft'' picture, assessing cognitive functions such as language comprehension and memory.

    \item \textbf{Phonemic Fluency (PFT).} Subjects utter as many words starting with the letter ``P'' as possible within a minute, evaluating executive function and phonological processing.

    \item \textbf{Semantic Fluency (SFT).} Subjects are asked to list as many animals as possible in one minute, assessing verbal fluency, lexical access, and semantic memory.
    
\end{itemize}

\begin{figure}[ht]
\centering
\includegraphics[width=0.75\columnwidth]{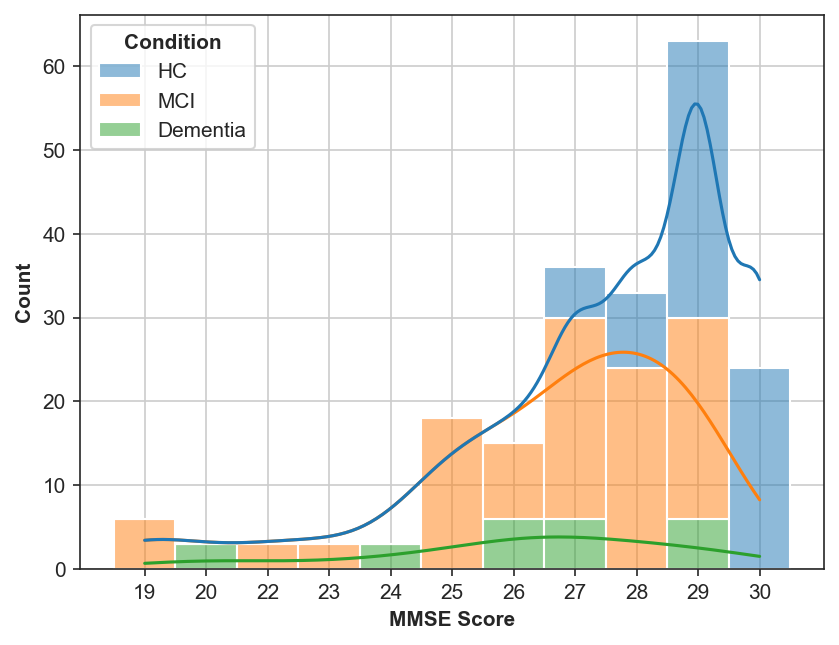}
\vspace{-6pt}
\caption{Distribution of the MMSE scores per diagnostic class.}
\label{fig:mmse}
\vspace{-6pt}
\end{figure}

\noindent\textbf{Challenges.} One of the main challenges in this dataset is the strong data imbalance observed between diagnostic classes, as shown in Table \ref{tab:demographics}. Although this reflects a real clinical scenario, the under-representation of the dementia class may affect model performance and generalization.
This is further compounded by the differences in the age distribution for dementia patients from the train to the development partition. 
Another challenge is the complexity and subjectivity inherent in clinical assessments. As illustrated in Figure \ref{fig:mmse}, the MMSE score distribution exhibits considerable overlap across conditions --- even between groups that are presumed to be more distinct, such as HC and Dementia --- highlighting the difficulty of differentiating between cognitive disorders in the early stages. Furthermore, only 69 speakers out of the 157 include annotations for MMSE scores. For these reasons, we concentrated our efforts on the classification task, rather than directly predicting the target MMSE scores via regression, as also proposed in the challenge.

\section{Method}
\label{sec:method}

This study evaluates various systems for distinguishing healthy controls, from individuals with MCI and individuals with Dementia, based on their speech recorded for the three tasks detailed in Section~\ref{sec:data}. The compared systems target the multiple individual manifestations of early onset of cognitive impairment, namely those observed in speech and language production in general, representing data based on some of the possible symptoms of this disease, as outlined below. The predictions from the best-performing systems were combined through late fusion, to integrate complementary insights from different systems to obtain a final prediction.

Early experiments revealed that many systems struggled to correctly classify dementia cases. Consequently, the selection of the best-performing systems prioritized a balanced macro F1-score across the training (via cross-validation) and development (held-out) sets, while ensuring an acceptable F1-score specifically for the dementia class.

Figure~\ref{fig:method} shows an overview of the proposed method, where speech signals are first preprocessed to remove noise prior to automatic transcription. Features are then extracted from both the cleaned speech signals and their corresponding transcripts, serving as inputs for classifier training. Finally, the best-performing systems are combined into a model ensemble to produce the final predictions.

\subsection{Pre-processing}

An initial manual analysis of the recordings identified two possible confounding factors: the presence of loud noises and the interventions of interviewers. While the former was dealt with through signal processing approaches, for the latter, we experimented with speaker diarization systems for the automatic removal of the interventions.
However, early experiments showed that state-of-the-art diarization systems had very poor performance, often removing large segments corresponding to the subject. For this reason, and after a preliminary evaluation with manual removal of external interventions that showed little difference in performance, we opted to skip this step.
As such, in what follows, we will provide short descriptions of the methods applied for noise removal and automatic transcription.

\vspace{6pt}
\noindent\textbf{Noise Removal.}
Preliminary examination of the data revealed that several recordings contained signaling tones from the recording protocol that marked the beginning and end of tasks, in addition to various other loud noises. To prevent such noises from interfering with downstream tasks, an energy-based de-noising step was introduced to replace these sounds with low-energy random Gaussian noise in unvoiced segments.

\vspace{6pt}
\noindent\textbf{Automatic Speech Recognition.} 
To automatically generate transcripts for speech recordings, we compared two versions of OpenAI's Whisper model\footnote{
The following prompt was provided to both whisper models: \textit{"The sentence may be cut off, do not make up words to fill in the rest of the sentence. Include repetitions, fillers, and disfluencies such as Umm, let me think like, hmm... Okay, here's what I'm, like, thinking."},~\cite{openai2024prompt}
} ~\cite{radford2023whisper}, \textit{whisper-medium.en} and \textit{whisper-large-v2}, as well as CrisperWhisper~\cite{zusag2024crisper}. CrisperWhisper is a variant of Whisper that aims to capture every spoken word exactly as it is, including disfluencies, fillers, pauses, and stutters. This characteristic makes CrisperWhisper a promising alternative to address the automatic identification of neurodegenerative disorders, given that disfluencies are often considered very informative.

Early analysis of the results revealed that CrisperWhisper captured more disfluencies than the Whisper models, often with spellings differing from the reference transcripts, leading to higher Word Error Rate (WER) penalties. To address this, we applied post-processing to standardize disfluencies by replacing them with a uniform placeholder, solely for the purpose of WER computation. Table~\ref{tab:asr} presents the performance of each model after this post-processing, categorized by speech task. 

Despite the hallucinations observed in the fluency tasks, CrisperWhisper reached lower WERs than its Whisper counterpart for the CTD and SFT tasks. Therefore, CrisperWhisper was used to extract automatic transcripts of CTD and SFT samples, while \textit{whisper-medium.en} was used for the PFT samples. 

A key takeaway from Table~\ref{tab:asr} is that fluency tasks appear to pose greater challenges for the evaluated ASR systems compared to CTD. This may stem from the fact that these systems are trained with continuous speech that observes grammatical rules, while these tasks correspond to listings with little grammatical structure, which highlights the limitations of these systems when applied to such clinical assessment tasks.

\begin{table}[t]
\caption{ASR performance in terms of WER (\%) per speech task.}
\vspace{-7pt}
\label{tab:asr}
\tiny
\centering
\setlength{\tabcolsep}{3pt}
\resizebox{0.7\linewidth}{!}{
\begin{tabular}{l c|ccc}
\toprule
& \textbf{All} & \textbf{CTD} & \textbf{PFT} &\textbf{SFT}  \\
\midrule
\textit{whisper-large-v2} & 35.9 & 31.9 & 49.7 & 41.3 \\
\textit{whisper-medium.en} & 26.3 & 19.3 & \textbf{43.7} & 39.6 \\
\textit{CrisperWhisper} & \textbf{23.2} & \textbf{14.1} & 54.7 & \textbf{34.8} \\
\bottomrule
\end{tabular}
}
\vspace{-10pt}
\end{table}

\subsection{Feature Extraction}

To model the manifestations of cognitive impairment, we used feature representations obtained from both the acoustic and linguistic components of speech, including neural and knowledge-based (KB) features.

\vspace{6pt}
\noindent\textbf{Acoustic features.}
In terms of KB features, we compared four sets of features: eGeMAPS~\cite{egemaps} and ComParE~\cite{compare}, extracted with OpenSMILE~\cite{eyben2010opensmile}; a set of 33 acoustic features extracted with \textit{Praat}~\cite{praat}, henceforth referred to as \textit{Praat}, and a subset of these, containing 11 rhythm-related features, referred to as \textit{Pauses}. The \textit{Praat} features, strongly overlap with those described in~\cite{botelho2024speech}.
In terms of neural speech representations, we considered the use of ECAPA-TDNN speaker embeddings (ECAPA)~\cite{desplanques2020ecapa} and TRILLsson paralinguistic embeddings~\cite{shor22trillsson}.

\vspace{6pt}
\noindent\textbf{Text features.}
In terms of KB features, we compared two feature sets: one set of 11 linguistic features (Ling.) previously used in~\cite{botelho2024speech}, and one set of 8 linguistic features designed for fluency tasks (Fluency).
Additionally, we explored the macrodescriptors (Macro.) proposed in~\cite{botelho2024macro}, estimated using Llama-3.1-70B-Instruct~\cite{llama3modelcard}. In terms of neural representations, we explored embeddings from pre-trained Language Models (PLMs), namely BERT \cite{devlin-etal-2019-bert}, RoBERTa \cite{liu2019roberta}, and LongFormer \cite{beltagy2020longformer}.

\subsection{Classification}

For the classification task, we use Support Vector Machines (SVMs), Decision Trees (DT), and Random Forest (RF). Besides, we explore Fuzzy Fingerprints (FFP), an interpretable classification technique that has proved to be useful when there is high class imbalance in text datasets \cite{pereira2023fuzzy}. 

To address data scarcity, the training set was divided into five folds, ensuring a balanced distribution of classes and gender. Classifiers were trained using five-fold cross-validation. The predictions for the development set were obtained from the five classifiers trained during cross-validation and were averaged to generate a single prediction before computing the final performance metrics.
All systems were evaluated in terms of unweighted average F1-score (UAF1), following the challenge guidelines. The F1-score for each class is also reported.
As the official test set was unavailable, performance is reported only for the training set (using cross-validation) and the held-out development set.

Most systems were trained for each of the three tasks; however, preliminary experiments showed that acoustic neural embedding-base systems benefited from a combination of the three tasks to obtain speaker-level predictions. 

\subsection{Model Ensemble}
\label{sec:ensemble}

After evaluating all single systems (i.e., one or multiple feature sets, followed by a classifier), we adopted a model ensemble strategy to exploit the potential complementary information embedded within the individual systems.

The top-performing systems were selected for ensemble modeling, with a focus on those exhibiting balanced performance across cross-validation and development sets, as well as non-zero F1-scores for the dementia class. Each system provided scores for each class -- either as probabilities, softmax scores, or, in the case of SVMs, distances to the class-separating hyperplane. These scores, or soft decisions, were aggregated using a multinomial logistic regression model, trained over the outputs obtained with the single systems for the training set during cross-validation.
The resulting model was then used to provide predictions for the development set. 
In these experiments, we considered all possible ensemble combinations of up to six single systems at a time.

\begin{figure*}[t]
    \centering
    \includegraphics[width=0.88\linewidth]{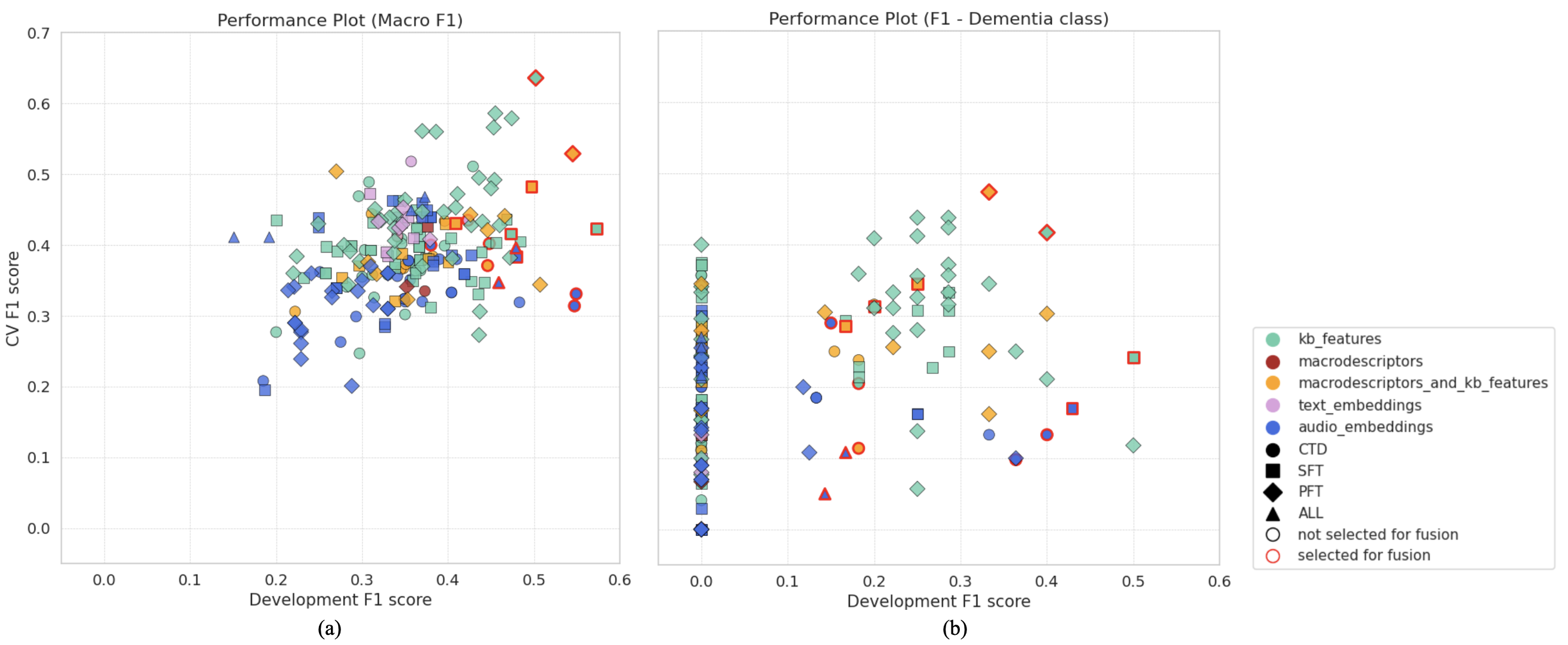}
    \caption{Single system performance, in terms of macro F1 (a) and F1 on the dementia class (b).}
    \vspace{-6pt}
    \label{fig:results_single_sys}
\end{figure*}
\begin{figure}[t]
    \centering
    \includegraphics[width=0.75\columnwidth]{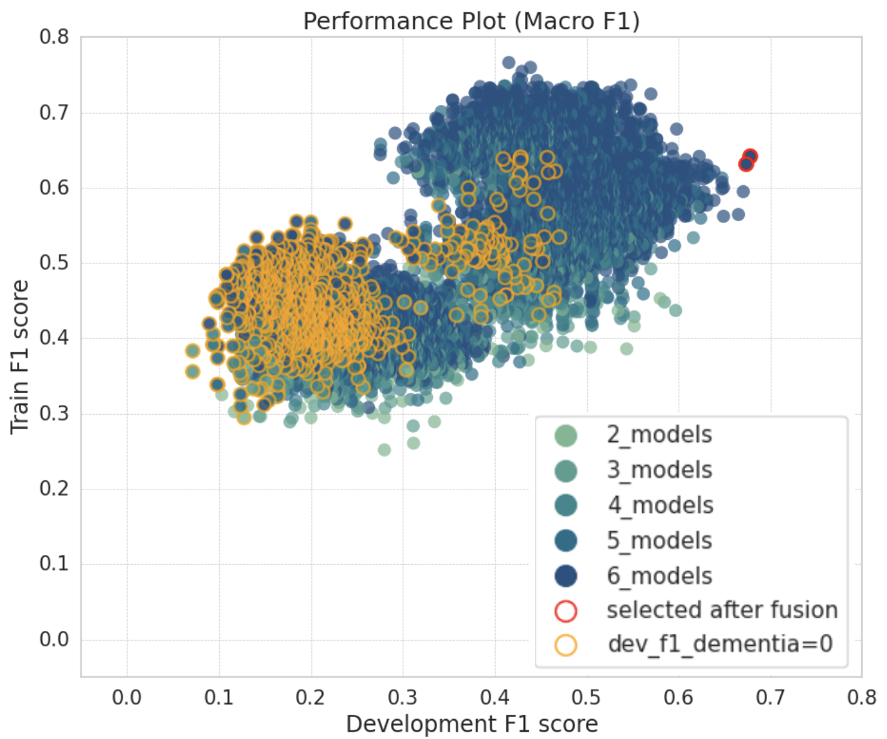}
    \caption{Model ensemble performance, in terms of macro F1.}
    \label{fig:results_ensemble}
    \vspace{-6pt}
\end{figure}

\section{Results \& Discussion}
\label{sec:results}

\noindent\textbf{Single-system experiments.} The results for the individual systems are represented in Figure~\ref{fig:results_single_sys}, where the best-performing models can be found in the top right corner. In this figure, we can observe how systems trained on the PFT task -- denoted by a $\diamond$ -- generally achieve the highest performance.
Moreover, KB features and their combination with macro-descriptors consistently yield superior results, when compared to other data representations.

Of the 205 systems evaluated, 15 were identified as promising candidates for fusion and subsequently considered in the model ensemble stage. These systems, marked with a red edge in Figure~\ref{fig:results_single_sys}, include both top- and medium-performing models. This diverse selection of approaches aimed to incorporate complementary information, potentially enhancing the overall effectiveness of the ensemble.

\noindent\textbf{Model ensemble experiments.} The resulting 15 models were then combined using the ensemble strategy described in Section \ref{sec:ensemble}. By defining groups of 2 up to 6 systems at a time, we explored around 10k system combinations in our experiments. 

The performance of these ensembled models is presented in Figure~\ref{fig:results_ensemble}. 
In this figure, it is possible to observe that a large number of systems are able to generalize their performance from the training set to the development set. Moreover, in general, ensembles including a larger number of single systems seem to correlate well with higher performance on both datasets, indicating that the proposed approaches provide complementary information. Furthermore, comparing these results with those of Figure \ref{fig:results_single_sys}, shows that model fusion through logistic regression substantially improves performance when compared with individual models. 
Nevertheless, we also observe that several models, highlighted in yellow, exhibited an F1 score of zero for the dementia class, showcasing the challenging nature of this dataset. One reason for this might be the age differences between the train and development partitions for the dementia subjects, as well as the significant data imbalance.

\begin{figure}[t]
\centering
\includegraphics[width=0.8\columnwidth]{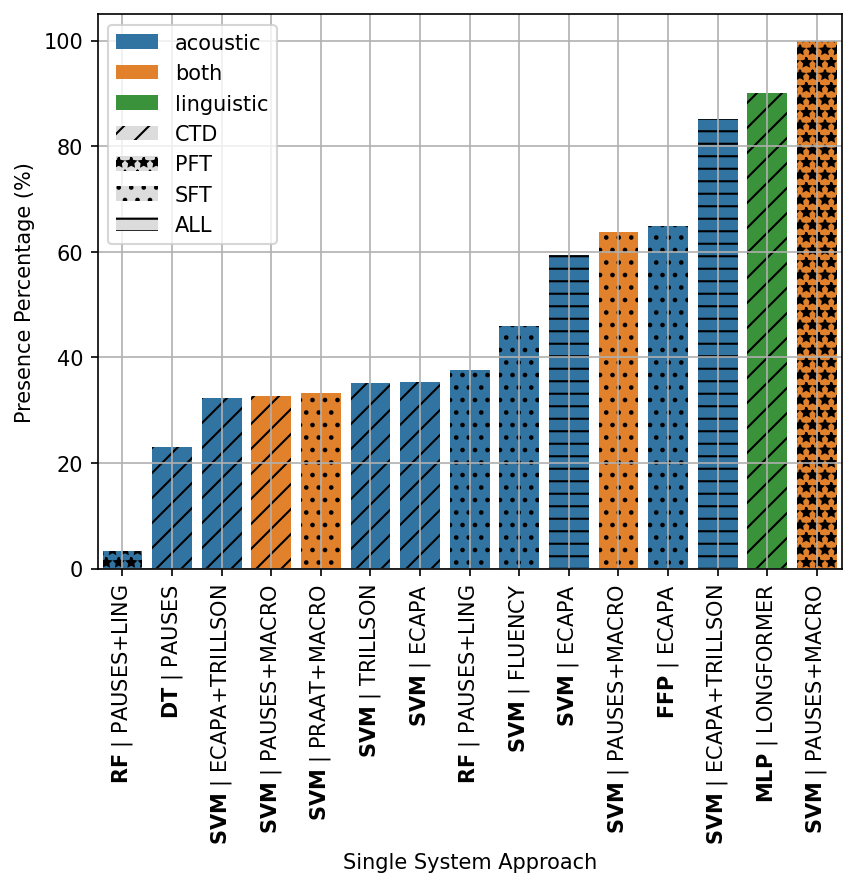}
\caption{Frequency of single classifiers appearing in the top best-performing ensemble model combinations, categorized by the type of features and tasks used during their training.}
\label{fig:freq}
\vspace{-10pt}
\end{figure}

\noindent\textbf{Fine-grained analysis.} For a more refined analysis, we selected 421 ensembles that achieve a macro F1 score above 55\% in the training and development sets and a non-zero F1 score in the dementia class. Figure~\ref{fig:freq} illustrates the frequency with which each of the 15 individual systems appears in this set of top-performing model ensembles. One key observation is that the system incorporating pause features and macrodescriptors trained on the PFT task is consistently selected, with the same system trained on SFT samples being selected over 60\% of the time. Additionally, the system based on the Longformer embeddings for the CTD task and the system based on the combination of ECAPA and TRILLsson embeddings for all tasks are selected in more than 80\% of the best performing model ensembles.

These observations show that both linguistic and acoustic data representations are very frequently selected, and thus should contain meaningful information. 
Not only neural representations, but also interpretable high-level concepts such as macrodescriptors and pause features are found to be very informative. In contrast, no clear patterns or trends were found in relation to the type of classifier and task used for their training.

\noindent\textbf{Final system selection.} As our final systems, we selected the two ensembles -- highlighted with a red edge in Figure~\ref{fig:results_ensemble} -- that offer the best balance between macro F1 in the train and development, and F1 score for dementia, henceforth referred to as \textit{Ensemble \#1} and \textit{Ensemble \#2}. Each ensemble consists of 6 single systems, with detailed descriptions provided in Table~\ref{tab:sys_in_model_ensemble}. Notably, 5 of the 6 systems are shared between the two ensembles. The results obtained with each of these systems, individually, as well as those obtained after fusion are presented in Table~\ref{tab:results}. 
The performance of the two fusion systems on the official test is 54.36\% and 59.34\%, for the \textit{Ensemble\#1} and \textit{Ensemble\#2}, respectively.

\begin{table*}[t]
\caption{Systems selected for the best performing model ensembles (Ens\#1 and Ens\#2).}
\vspace{-7pt}
\label{tab:sys_in_model_ensemble}
\centering
\setlength{\tabcolsep}{3pt}
\resizebox{\linewidth}{!}{
\begin{tabular}{c l l l c c p{14cm}}
\toprule
\textbf{System \#} & \textbf{Data representation} & \textbf{Task} & \textbf{Classifier} & \textbf{Ens\#1} & \textbf{Ens\#2} & \textbf{Description}  \\
\midrule
\textbf{1} & Longformer & CTD & MLP & -- & \checkmark & This system corresponds to a finetuning of the pre-trained LongFormer~\cite{beltagy2020longformer}. The Longformer uses modified attention mechanisms, acting on both local and global scale, which allows for tackling longer texts. We feed the input transcriptions to the LongFormer, use the $[CLS]$ token as pooling strategy and resort to a fully connected linear layer for classification. \\

\noalign{\vskip 0.5ex} \hdashline \noalign{\vskip 0.5ex}

\textbf{2} & Fluency & SFT & SVM & \checkmark & \checkmark & Prior to extracting fluency features for SFT, transcripts were processed an LLM, specifically Llama-3.1-70B-Instruct, prompted to identify all "target words" corresponding to animals. Six features were then computed, including counts and repetitions of unique target words, as well as the mean and standard deviation of cosine similarities between Word2Vec embeddings of adjacent target words. These features were input into an SVM with a linear kernel and C=$0.1$. \\
 
\noalign{\vskip 0.5ex} \hdashline \noalign{\vskip 0.5ex}

\textbf{3} & Pauses + Macro & PFT  & SVM & \checkmark & \checkmark & The set of 11 pause related features, and four macrodescriptors were fed to an SVM with a linear kernel and C=$0.1$, resulting in two systems, one for PFT and another for SFT. \\

\textbf{4} & & SFT & SVM & \checkmark & \checkmark & \\

\noalign{\vskip 0.5ex} \hdashline \noalign{\vskip 0.5ex}

\textbf{5} & ECAPA & SFT & FFP & \checkmark & -- & This system leverages ECAPA-TDNN embeddings extracted from the SFT task, using Fuzzy Fingerprints as a classification strategy. \\

\noalign{\vskip 0.5ex} \hdashline \noalign{\vskip 0.5ex}

\textbf{6} & ECAPA & All & SVM & \checkmark & \checkmark & This system corresponds to a linear SVM, C=$0.0001$, trained over the concatenation of ECAPA-TDNN embeddings, for all three tasks. PCA dimensionality reduction was applied to the resulting vector. \\

\textbf{7} & ECAPA + TRILLsson & All & SVM & \checkmark & \checkmark & Equal to the system above, however, in this case the PCA-reduced set of ECAPA-TDNN embeddings is concatenated with the set of PCA-reducted TRILLsson embeddings. \\

\bottomrule
\end{tabular}
}
\end{table*}
\begin{table}[t]
\caption{Results for the best-performing model ensembles as well as for the individual systems that were included in the best-performing model ensembles. Results are presented in \%.}
\vspace{-7pt}
\label{tab:results}
\centering
\setlength{\tabcolsep}{3pt} 
\resizebox{\linewidth}{!}{
\begin{tabular}{lcccccccc}

\toprule
& & \multicolumn{3}{c}{\textbf{Cross-validation}} & \multicolumn{4}{c}{\textbf{Held-out Development}} \\ 
\textbf{System \#} & \textbf{UAF\textsubscript{1}} & F\textsubscript{1 HC} & F\textsubscript{1 MCI} & F\textsubscript{1 Dem} & UAF\textsubscript{1} & F\textsubscript{1 HC} & F\textsubscript{1 MCI} & F\textsubscript{1 Dem}  \\
\midrule
\multicolumn{8}{l}{\textbf{Single system}} \\
\midrule
\textbf{1} & 43.5 & 72.8 & 57.7 & 0.0 & 42.3 & 73.3 & 53.6 & 0.0 \\
\textbf{2} & 42.3 & 60.3 & 42.4 & 24.2 & 57.3 & 64.9 & 57.1 & 50.0  \\
\textbf{3} & 43.1 & 64.9 & 35.9 & 28.6 & 40.9 & 72.7 & 33.3 & 16.7  \\
\textbf{4} & 52.9 & 58.9 & 52.4 & 47.4 & 54.5 & 63.4 & 66.7 & 33.3 \\
\textbf{5} & 38.4 & 60.4 & 37.8 & 17.0 & 47.9 & 57.9 & 42.9 & 42.9  \\
\textbf{6} & 34.7 & 59.0 & 40.0 & 5.0 & 45.9 & 73.7 & 50.0 & 14.3 \\
\textbf{7} & 39.6 & 63.6 & 44.4 & 10.8 & 47.9 & 73.7 & 53.3 & 16.7 \\
\midrule
\multicolumn{8}{l}{\textbf{Model ensemble}} \\
\midrule
Ensemble \#1 & 64.2 & 76.9 & 62.7 & 52.9 & 67.8 & 77.3 & 69.0 & 57.2 \\
Ensemble \#2 & 63.1 & 75.6 & 58.2 & 55.6 & 67.4 & 78.3 & 66.7 & 57.1 \\

\bottomrule
\end{tabular}
}
\vspace{-10pt}
\end{table}

\section{Limitations}
\label{sec:limitations}

Among the limitations of this work, we first highlight the challenges arising from the dataset, including class imbalance and overlapping MMSE scores between classes -- factors which may partly explain the poorer performance compared to prior studies on ADReSS~\cite{yuan2020adress_best, botelho2024macro} or Taukadial~\cite{perez2024multilingual} challenges, even considering that those were two-class classification problems.

A second limitation is that our approach does not consider demographic data, which physicians typically consider in clinical scenarios. This omission is due to two factors: the dataset's limited size and the unavailability of demographic information on the released test set, making its incorporation infeasible in our approach. This also has a negative impact in the "real-word" applicability of our approach and dataset in general, since key demographic information is crucial for accurate clinical-decision making.

A third limitation is the potential influence of interviewer interventions on the results. An automated diagnostic tool should not rely on the behavior of healthcare practitioners. While this issue could be partially mitigated through diarization, our experiments with state-of-the-art diarizers failed to yield high-quality outputs. Moreover, even with diarization, the subject's behavior remains influenced by interventions, such as prompts to recall specific information (e.g., "farm animals"). While it is natural for individuals with cognitive impairment to require additional support during speech tasks, this dependency highlights a key limitation of the approach.

Finally, the potential for overfitting must be acknowledged. Despite using cross-validation on the training set and maintaining a held-out development set, the risk of overfitting increases due to the extensive comparison of approaches on a limited dataset. Future research and validation on larger datasets are necessary to address this issue. 

\section{Conclusions}
\label{sec:conclusion}

This paper presents our joint group efforts for the PROCESS challenge \cite{christensen2024process}, which aims to assess cognitive decline through the automatic analysis of spontaneous speech production. We conducted extensive experimentation, addressing the challenge from multiple perspectives. This included evaluating the accuracy of various ASR models, exploring numerous acoustic- and linguistic-based feature descriptors -- as well as their combinations -- and studying a wide range of classification approaches. Our findings suggest that model ensembles of diverse systems achieve higher performance.
Further analyses show how the SVM-based systems using pause-related speech features and macro descriptors \cite{botelho2024macro} are consistently selected across model ensembles, highlighting the effectiveness of KB speech features and high-level assessment concepts as potential biomarkers of cognitive decline.
Additional insights demonstrate the capability of utterance-level speech embeddings to encode and fuse information from multiple speech assessment tasks, as well as of Longformer \cite{beltagy2020longformer} in capturing long-term dependencies, particularly in the CTD task, where context processing is essential. 
The limitations encountered, however, highlight the ongoing research challenges in developing automatic systems for cognitive decline assessment, particularly in real-world scenarios involving transitional stages and significant data imbalance.

\section{Acknowledgements}
The work of Gimeno-Gómez and Martínez-Hinarejos was partially supported by GVA through Grants CIACIF/2021/295 and CIBEFP/2023/167, by Grant PID2021-124719OB-I00 under project LLEER funded by MCIN/AEI and ERDF, EU ``A way of making Europe". The work of the remaining authors was supported by Portuguese national funds through Fundação para a Ciência e a Tecnologia, with reference DOI: 10.54499/UIDB/50021/2020 and reference UI/BD/154561/2022, and in part by Portuguese Recovery and Resilience Plan and NextGenerationEU European Union Funds under Project C644865762-00000008 (Accelerat.AI) and C645008882-00000055 (Responsible.AI). We extend our special thanks to the organizers of the challenge.

\bibliographystyle{IEEEtran}
\bibliography{main}

\begin{thebibliography}{10}
\providecommand{\url}[1]{#1}
\csname url@samestyle\endcsname
\providecommand{\newblock}{\relax}
\providecommand{\bibinfo}[2]{#2}
\providecommand{\BIBentrySTDinterwordspacing}{\spaceskip=0pt\relax}
\providecommand{\BIBentryALTinterwordstretchfactor}{4}
\providecommand{\BIBentryALTinterwordspacing}{\spaceskip=\fontdimen2\font plus
\BIBentryALTinterwordstretchfactor\fontdimen3\font minus \fontdimen4\font\relax}
\providecommand{\BIBforeignlanguage}[2]{{%
\expandafter\ifx\csname l@#1\endcsname\relax
\typeout{** WARNING: IEEEtran.bst: No hyphenation pattern has been}%
\typeout{** loaded for the language `#1'. Using the pattern for}%
\typeout{** the default language instead.}%
\else
\language=\csname l@#1\endcsname
\fi
#2}}
\providecommand{\BIBdecl}{\relax}
\BIBdecl

\bibitem{who-dementia}
W.~H. Organization, ``Dementia,'' \url{https://www.who.int/news-room/fact-sheets/detail/dementia}, accessed on September 28, 2021.

\bibitem{snowdon1996linguistic}
D.~A. Snowdon, S.~J. Kemper, J.~A. Mortimer, L.~H. Greiner, D.~R. Wekstein, and W.~R. Markesbery, ``Linguistic ability in early life and cognitive function and alzheimer's disease in late life: Findings from the nun study,'' \emph{Jama}, vol. 275, no.~7, pp. 528--532, 1996.

\bibitem{kempler1995}
D.~Kempler, ``{Language changes in dementia of the Alzheimer type},'' \emph{Dementia and communication}, pp. 98--114, 1995.

\bibitem{reilly2011}
J.~Reilly, J.~Troche, and M.~Grossman, ``{Language processing in dementia},'' \emph{{The handbook of Alzheimer's Disease and Other Dementias}}, pp. 336--368, 2011.

\bibitem{petersen2001current}
R.~C. Petersen, R.~Doody, A.~Kurz, R.~C. Mohs, J.~C. Morris, P.~V. Rabins, K.~Ritchie, M.~Rossor, L.~Thal, and B.~Winblad, ``Current concepts in mild cognitive impairment,'' \emph{Archives of neurology}, vol.~58, no.~12, pp. 1985--1992, 2001.

\bibitem{who2017}
\BIBentryALTinterwordspacing
W.~H. Organization, ``Evidence profile: Cognitive impairment,'' 2017. [Online]. Available: \url{https://iris.who.int/bitstream/handle/10665/342246/WHO-MCA-17.06.02-eng.pdf}
\BIBentrySTDinterwordspacing

\bibitem{satt2014speech}
A.~Satt, R.~Hoory, A.~K{\"o}nig, P.~Aalten, and P.~H. Robert, ``Speech-based automatic and robust detection of very early dementia,'' in \emph{Fifteenth Annual Conference of the International Speech Communication Association}, 2014.

\bibitem{bertini2021automatic}
F.~Bertini, D.~Allevi, G.~Lutero, D.~Montesi, and L.~Calz{\`a}, ``Automatic speech classifier for mild cognitive impairment and early dementia,'' \emph{ACM Transactions on Computing for Healthcare (HEALTH)}, vol.~3, no.~1, pp. 1--11, 2021.

\bibitem{perez2024multilingual}
P.~A. P{\'e}rez-Toro, T.~Arias-Vergara, P.~Klumpp, T.~Weise, M.~Schuster, E.~Noeth, J.~R. Orozco-Arroyave, and A.~Maier, ``Multilingual speech and language analysis for the assessment of mild cognitive impairment: Outcomes from the taukadial challenge,'' \emph{Proc. Interspeech 2024}, pp. 982--986, 2024.

\bibitem{pompili2020pragmatic}
A.~Pompili, A.~Abad, D.~M. de~Matos, and I.~P. Martins, ``Pragmatic aspects of discourse production for the automatic identification of alzheimer's disease,'' \emph{IEEE Journal of Selected Topics in Signal Processing}, vol.~14, no.~2, pp. 261--271, 2020.

\bibitem{ablimit_botelho2022exploring}
A.~Ablimit, C.~Botelho, A.~Abad, T.~Schultz, and I.~Trancoso, ``Exploring dementia detection from speech: Cross corpus analysis,'' in \emph{ICASSP 2022-2022 IEEE International Conference on Acoustics, Speech and Signal Processing (ICASSP)}.\hskip 1em plus 0.5em minus 0.4em\relax IEEE, 2022, pp. 6472--6476.

\bibitem{luz2021alzheimer}
S.~Luz, F.~Haider, S.~de~la Fuente~Garcia, D.~Fromm, and B.~MacWhinney, ``Alzheimer's dementia recognition through spontaneous speech,'' \emph{Frontiers in computer science}, vol.~3, p. 780169, 2021.

\bibitem{ADReSS}
S.~Luz, F.~Haider, S.~de~la Fuente, D.~Fromm, and B.~MacWhinney, ``{Alzheimer’s Dementia Recognition Through Spontaneous Speech: The ADReSS Challenge},'' in \emph{Proc. Interspeech 2020}, 2020, pp. 2172--2176.

\bibitem{ADReSSo_luz2021detecting}
------, ``Detecting cognitive decline using speech only: The {ADReSSo} challenge,'' in \emph{INTERSPEECH 2021}.\hskip 1em plus 0.5em minus 0.4em\relax ISCA, 2021.

\bibitem{ADReSSM_luz2023multilingual}
S.~Luz, F.~Haider, D.~Fromm, I.~Lazarou, I.~Kompatsiaris, and B.~MacWhinney, ``Multilingual alzheimer’s dementia recognition through spontaneous speech: a signal processing grand challenge,'' in \emph{ICASSP 2023-2023 IEEE International Conference on Acoustics, Speech and Signal Processing (ICASSP)}.\hskip 1em plus 0.5em minus 0.4em\relax IEEE, 2023, pp. 1--2.

\bibitem{TAUKADIAL_luz2024connected}
S.~Luz, S.~D. L.~F. Garcia, F.~Haider, D.~Fromm, B.~MacWhinney, A.~Lanzi, Y.-N. Chang, C.-J. Chou, and Y.-C. Liu, ``Connected speech-based cognitive assessment in chinese and english,'' \emph{arXiv preprint arXiv:2406.10272}, 2024.

\bibitem{christensen2024process}
H.~Christensen, S.~Bell, D.~Blackburn, B.~Mirheidari, M.~Pahar, F.~Tao, D.~Braun, H.~ElGhazaly, C.~Illingworth, O.~Ronan, F.~Peters, S.~Luz, and F.~Haider, ``{ICASSP 2025 SPGC Challenge: PROCESS},'' \url{https://processchallenge.github.io/}, 2024, [Online; accessed 30-December-2024].

\bibitem{openai2024prompt}
OpenAI, ``{OpenAI Platform Speech to Text: Prompting},'' \url{https://platform.openai.com/docs/guides/speech-to-text/prompting}, 2024, [Online; accessed 30-December-2024].

\bibitem{radford2023whisper}
A.~Radford, J.~W. Kim, T.~Xu, G.~Brockman, C.~McLeavey, and I.~Sutskever, ``{Robust speech recognition via large-scale weak supervision},'' in \emph{ICML}.\hskip 1em plus 0.5em minus 0.4em\relax PMLR, 2023, pp. 28\,492--28\,518.

\bibitem{zusag2024crisper}
M.~Zusag, L.~Wagner, and B.~Thallinger, ``{CrisperWhisper: Accurate Timestamps on Verbatim Speech Transcriptions},'' in \emph{Interspeech}, 2024, pp. 1265--1269.

\bibitem{egemaps}
F.~Eyben, K.~Scherer, B.~Schuller, J.~Sundberg, E.~André, C.~Busso, L.~Devillers, J.~Epps, P.~Laukka, S.~Narayanan, and K.~Truong, ``{The Geneva Minimalistic Acoustic Parameter Set (GeMAPS) for Voice Research and Affective Computing},'' \emph{IEEE transactions on affective computing}, vol.~7, no.~2, pp. 190--202, 4 2016, open access.

\bibitem{compare}
B.~Schuller, S.~Steidl, A.~Batliner, A.~Vinciarelli, K.~R. Scherer, F.~Ringeval, M.~Chetouani, F.~Weninger, F.~Eyben, E.~Marchi \emph{et~al.}, ``{The INTERSPEECH 2013 Computational Paralinguistics Challenge: Social Signals, Conflict, Emotion, Autism},'' in \emph{INTERSPEECH 2013: 14th Annual Conference of the International Speech Communication Association}, 2013.

\bibitem{eyben2010opensmile}
F.~Eyben, M.~W{\"o}llmer, and B.~Schuller, ``{OpenSMILE}: the munich versatile and fast open-source audio feature extractor,'' in \emph{Proceedings of the 18th ACM international conference on Multimedia}, 2010, pp. 1459--1462.

\bibitem{praat}
\BIBentryALTinterwordspacing
P.~Boersma and D.~Weenink, ``Praat: doing phonetics by computer [computer program].'' [Online]. Available: \url{http://www.praat.org/}
\BIBentrySTDinterwordspacing

\bibitem{botelho2024speech}
C.~Botelho, A.~Abad, T.~Schultz, and I.~Trancoso, ``Speech as a biomarker for disease detection,'' \emph{IEEE Access}, 2024.

\bibitem{desplanques2020ecapa}
B.~Desplanques, J.~Thienpondt, and K.~Demuynck, ``{ECAPA-TDNN: Emphasized Channel Attention, Propagation and Aggregation in TDNN Based Speaker Verification},'' in \emph{Proc. Interspeech}, 2020, pp. 3830--3834.

\bibitem{shor22trillsson}
J.~Shor and S.~Venugopalan, ``{TRILLsson: Distilled Universal Paralinguistic Speech Representations},'' in \emph{Interspeech}, 2022, pp. 356--360.

\bibitem{botelho2024macro}
C.~Botelho, J.~Mendon{\c{c}}a, A.~Pompili, T.~Schultz, A.~Abad, and I.~Trancoso, ``Macro-descriptors for alzheimer’s disease detection using large language models,'' in \emph{Proc. Interspeech}, 2024, pp. 1975--1979.

\bibitem{llama3modelcard}
\BIBentryALTinterwordspacing
AI@Meta, ``Llama 3 model card,'' 2024. [Online]. Available: \url{https://github.com/meta-llama/llama3/blob/main/MODEL_CARD.md}
\BIBentrySTDinterwordspacing

\bibitem{devlin-etal-2019-bert}
J.~Devlin, M.-W. Chang, K.~Lee, and K.~Toutanova, ``{BERT}: Pre-training of deep bidirectional transformers for language understanding,'' in \emph{Proceedings of the 2019 Conference of the North {A}merican Chapter of the Association for Computational Linguistics: Human Language Technologies, Volume 1 (Long and Short Papers)}.\hskip 1em plus 0.5em minus 0.4em\relax Minneapolis, Minnesota: Association for Computational Linguistics, Jun. 2019, pp. 4171--4186.

\bibitem{liu2019roberta}
Y.~Liu, M.~Ott, N.~Goyal, J.~Du, M.~Joshi, D.~Chen, O.~Levy, M.~Lewis, L.~Zettlemoyer, and V.~Stoyanov, ``Roberta: A robustly optimized bert pretraining approach,'' \emph{arXiv preprint arXiv:1907.11692}, 2019.

\bibitem{beltagy2020longformer}
I.~Beltagy, M.~E. Peters, and A.~Cohan, ``Longformer: The long-document transformer,'' \emph{arXiv preprint arXiv:2004.05150}, 2020.

\bibitem{pereira2023fuzzy}
P.~Pereira, R.~Ribeiro, H.~Moniz, L.~Coheur, and J.~P. Carvalho, ``Fuzzy fingerprinting transformer language-models for emotion recognition in conversations,'' in \emph{2023 IEEE International Conference on Fuzzy Systems (FUZZ)}.\hskip 1em plus 0.5em minus 0.4em\relax IEEE, 2023, pp. 1--6.

\bibitem{yuan2020adress_best}
J.~Yuan, Y.~Bian, X.~Cai, J.~Huang, Z.~Ye, and K.~Church, ``Disfluencies and fine-tuning pre-trained language models for detection of {A}lzheimer's disease.'' in \emph{Interspeech}, 2020.

\end{thebibliography}

\end{document}